\documentclass[preprint,aps,showpacs]{revtex4}

\newcommand {\bea}{\begin{eqnarray}}
\newcommand {\eea}{\end{eqnarray}}
\newcommand {\be}{\begin{equation}}
\newcommand {\ee}{\end{equation}}
\newcommand {\pslash}{p\!\!\!/}
\usepackage{graphicx}
\usepackage{epsfig,subfigure}
\usepackage{amsmath}
\begin{document}

\def\Journal#1#2#3#4{{#1} {\bf #2}, #3 (#4)}
\def\RPP{{Rep. Prog. Phys.}}
\def\PRC{{Phys. Rev. C.}}
\def\PRD{{Phys. Rev. D.}}
\def\PR{{Phys. Rev.}}
\def\FP{{Foundations of Physics}}
\def\ZPA{{Z. Phys.A.}}
\def\NPA{{Nucl. Phys. A.}}
\def\JPG{{J. Phys. G Nucl. Part}}
\def\PRL{{Phys. Rev. Lett.}}
\def\PRpt{{Phys. Rep.}}
\def\PLB{{Phys. Lett. B}}
\def\AP{{Ann. Phys (N.Y.)}}
\def\EPJA{{Eur. Phys. J.A}}
\def\NP{{Nucl. Phys}}
\def\ZP{{Z. Phys}}
\def\RMP{{Rev. Mod. Phys}}
\def\IJMPE{{Int. J. Mod. Phys. E}}
\input epsf

  \title{Isovector Channel Role of Relativistic Mean Field Models in the Neutrino Mean Free Path}

\author{A. Sulaksono, P.T.P. Hutauruk, T. Mart}

\affiliation{Departemen Fisika, FMIPA, Universitas Indonesia,
Depok, 16424, Indonesia}

\begin{abstract}
An improvement in the treatment of the isovector channel of relativistic mean field (RMF) models based on effective field theory (E-RMF) is suggested, by adding an isovector scalar ($\delta$) meson and using a similar procedure to the one used by Horowitz and Piekarewicz to adjust the isovector-vector channel in order to achieve a softer density dependent symmetry energy of the nuclear matter at high density. Their effects on the equation of state (EOS) at high density and on the neutrino mean free path (NMFP) in neutron stars are discussed.
\end{abstract}

\pacs{13.15.+g, 25.30.Pt, 97.60.Jd}
\maketitle

\newpage

\section{Introduction}
\label{sec_intro}
Neutrino transport in stellar matter plays an important role in some phenomena such as the mechanism of supernovae explosions, structure of protoneutron stars, etc. Theoretical input needed for understanding the neutrino transport is the neutrino mean free path (NMFP) and the equation of state (EOS). A unified matter  model used in calculating  both observables is a requirement for having a  consistent neutrino transport result. We note that considerable efforts have been devoted to investigate the neutrino interaction in matter at high density~\cite{reddy1,niembro01,parada04,horo1,mornas1,mornas2,reddy2,horowitz91,yama,caiwan,margue,chandra,cowel,caroline05} with different matter models, variety of approximation  levels and purposes.

To this end, a kind of matter models that can be used is the relativistic mean field (RMF) model. The original or standard RMF models use the sigma, omega, and rho mesons with  additional cubic and quartic nonlinearities of sigma meson to effectively describe the interaction among nucleons. NL-3 parameter set~\cite{lala} belongs to this model. The parameter set has been very successful in the description of a variety of ground state properties of nuclei~\cite{todd}. Details of the models as well as their applications can be seen in Refs.~\cite{pg,ring,serot}. The effects of including delta meson in standard RMF models  including the corresponding linear responses have been studied for the asymmetric nuclear matter at low densities in Refs.~\cite{kubis97,liu02,greco03}, whereas for heavy ion collisions in Refs.~\cite{greco2,gaitanos1,gaitanos2} and on the neutron star properties in Ref.~\cite{liu04}. It was found that the $\delta$ field leads to a larger repulsion in the dense neutron rich matter (stiffer symmetry energy that leads to a larger proton fraction at high density), as well as a definite splitting of proton and neutron effective masses. Both features are influencing the stability conditions of a neutron star~\cite{liu04}. 

Inspired by the concept of effective field and density functional theories for hadrons, Furnstahl, Serot and Tang~\cite{Furnstahl96} constructed a new RMF model (from now on, will be denoted by E-RMF) that can be considered as an extension of the standard RMF models. One of the parameter sets in this model is G2. Besides yielding accurate predictions in finite nuclei and normal nuclear matter~\cite{serot,Furnstahl96,sil}, G2 has the interesting features like a positive value of quartic sigma meson coupling constant that leads to the existing lower bound in energy spectrum of this model~\cite{arumu,baym} and to the missing zero sound mode in the high density symmetric nuclear matter~\cite{cailon}. Moreover, the agreement of the nuclear matter and the neutron matter EOS at high density of G2 with the Dirac Brueckner Hartree Fock (DBHF) calculation ~\cite{sil,arumu} is better than those of NL-3, NL-1 and TM1 models (the standard RMF plus a quartic omega meson interaction). Nevertheless, from  the comparison between the neutron matter EOS of this model and that of the DBHF  result, the authors of Ref.~\cite{sil} pointed out that the present form of the E-RMF model still needs a substantial  improvement in the treatment of the isovector sector. It has also been shown  that the G2 parameters set of the E-RMF model still predicts a too large proton fraction~\cite{parada04}.  It is known that proton fraction correlates to the direct URCA cooling process~\cite{lati}. It is also known that this problem is caused by the role of isovector terms.  So far, the effects of the  delta meson inclusion on this model has not been studied yet.

Therefore, in this work, first, we will study the effects of adding an isovector-scalar ($\delta$) meson in the E-RMF model and the effects of the  isovector-vector channel adjustment by using a similar procedure to the one used in Ref.~\cite{Horowitz01}. The aim of these adjustments is to achieve a softer density dependent symmetry energy of nuclear matter at high density. The symmetry energy has a wide range of effects, such as  from giant dipole resonances to heavy ion collisions in nuclear physics and from supernovae to neutron stars  in astrophysics. In spite of its diverse influences, its magnitude and density dependence are not well understood~\cite{steiner}. More detail informations on the role of symmetry energy and related topics can be consulted to Ref.~\cite{steiner} and references therein. Second, we will also extend the analysis of our previous report~\cite{parada04} to give a more solid argument about the source of the possible appearance of the anomalous behavior in the NMFP of neutron stars predicted by RMF models. Here the anomalous behavior in the NMFP  means a contra intuitive NMFP results in a form of the decreasing of the mean free path with respect to the decreasing of the matter density~\cite{niembro01}. It has been known that this anomalous behavior exists in the NMFP predicted by non relativistic nuclear models of the Skyrme type~\cite{reddy1,niembro01}. This  anomalous behavior is attributed to the dominating term at high density that responsible to the appearance of the acausal behavior (the speed of sound exceeds the speed of light) of the model at high densities~\cite{reddy1}. It has been reported in Refs.~\cite{reddy1,reddy2,niembro01} that relativistic models alleviate this problem. But we have eventually found that not every parameter set of relativistic models is free from this problem~\cite{parada04}. 

In section~\ref{sec_frml} we will briefly explain the  formalism used in this work. In section ~\ref{sec_rslt} we discuss the results of our calculations. We will give the summary of our findings in section~\ref{sec_sum}. 

\section{Formalism}
This section contains a very brief description of the self consistent models for  nucleons plus the standard non-interacting Lagrangian densities for electron and muon as well as the interaction between neutrino-electron  with matter based on the standard model of weak interaction.
Here, a similar assumption with that of Refs.~\cite{niembro01,parada04} is used, i.e., the ground state of the neutron star is reached once the temperature has decreased below a few MeV. This state is gradually reached from the later stages of the cooling phase. The system is then quite dense and cool so that the zero temperature approximation is valid. In this case the direct URCA neutrino-neutron scattering is kinematically possible for low energy neutrinos at and above the threshold density when the proton fraction exceeds 1/9~\cite{lati} or slightly larger if muons are present. Furthermore, the absorption reaction is suppressed. For simplicity, we neglect the RPA correlations.
\label{sec_frml}
\subsection{E-RMF Model}
The calculation is done in the framework of the relativistic mean field approximation. The effective Lagrangian density used to describe nucleons interactions is taken from Refs.~\cite{Furnstahl96, Wang00, sil}. This Lagrangian is constructed with a nonlinear realization of chiral symmetry~\cite{Furnstahl96}. The explicit form of the E-RMF effective Lagrangian density reads
\be
 {\mathcal L}^{\rm nuc}= {\mathcal L}_N + {\mathcal L}_M . 
\ee

For nucleons, the Lagrangian density up to order $\nu=3$ is given by 
\bea
{\mathcal L}_N &=&\bar{\psi}[i \gamma^{\mu}(\partial_{\mu}+i \bar{\nu}_{\mu}+i g_{\rho}  \bar{b}_{\mu}+i g_{\omega} V_{\mu})+g_A  \gamma^{\mu} \gamma^{5} \bar{a}_{\mu}-M\nonumber\\&+&g_{\sigma} \sigma]\psi-\frac{f_{\rho} g_{\rho}}{4 M }\bar{\psi}\bar{b}_{\mu \nu} \sigma^{\mu \nu}\psi,
\eea
where
\be
\psi=\left( {p \atop n}\right),  ~ ~ ~ ~\bar{\nu}_{\mu}=-\frac{i}{2}(\bar{\xi}^{\dagger}\partial_{\mu}\bar{\xi}+\bar{\xi}\partial_{\mu}\bar{\xi}^{\dagger})= \bar{\nu}_{\mu}^{\dagger},
\ee
\be
\bar{a}_{\mu}=-\frac{i}{2}(\bar{\xi}^{\dagger}\partial_{\mu}\bar{\xi}-\bar{\xi}\partial_{\mu}\bar{\xi}^{\dagger})= \bar{a}_{\mu}^{\dagger},
\ee
\be
\bar{\xi}= {\rm exp}(i \bar{\pi}(x)/f_{\pi}), ~ ~ ~ ~ \bar{\pi}(x)=\frac{1}{2} \vec{\tau}\cdot \vec{\pi}(x),
\ee
\be
\bar{\pi}(x)=\frac{1}{2} \vec{\tau}\cdot \vec{\pi}(x),
\ee
\be
\bar{b}_{\mu \nu} = D_{\mu}\bar{b}_{\nu}-D_{\nu}\bar{b}_{\mu}+i g_{\rho}[\bar{b}_{\mu},\bar{b}_{\nu}],  ~ ~ ~ ~  D_{\mu}=\partial_{\mu}+i\bar{\nu}_{\mu},
\ee
\be
V_{\mu \nu}=\partial_{\mu}V_{\nu}-\partial_{\nu}V_{\mu},
\ee
\be
\sigma^{\mu \nu}=\frac{1}{2}[\gamma^{\mu},\gamma^{\nu}].
\ee
Here, $\it{p}$,  $\it{n}$  and  $\it{M}$ are the proton-, neutron-field and nucleon mass,  while $\sigma$, $\vec{\pi}$, $V^{\mu}$, and $\vec{b}^{\mu}$ are the $\sigma$,  $\pi$, $\omega$ and $\rho$ meson fields, respectively.
For mesons, the Lagrangian density up to order $\nu=4$ reads
\bea
{\mathcal L}_M &=&\frac{1}{4}f_{\pi}^2 {\rm Tr} (\partial_{\mu}\bar{U}\partial^{\mu}\bar{U}^{\dagger})+\frac{1}{4}f_{\pi}^2 {\rm Tr}(\bar{U} \bar{U}^{\dagger}-2)+\frac{1}{2}\partial_{\mu}\sigma\partial^{\mu}\sigma -\frac{1}{2} {\rm Tr}(\bar{b}_{\mu \nu}\bar{b}^{\mu \nu})-\frac{1}{4}V_{\mu \nu}V^{\mu \nu}\nonumber\\&-&g_{\rho \pi \pi}\frac{2 f_{\pi}^2}{m_{\rho}^2} {\rm Tr}(\bar{b}_{\mu \nu}\bar{\nu}^{\mu \nu})
+\frac{1}{2}(1+\eta_1 \frac{g_{\sigma} \sigma}{M}+\frac{\eta_2}{2} \frac{g_{\sigma}^2 \sigma^2}{M^2})m_{\omega}^2 V_{\mu}V^{\mu}+\frac{1}{4!}\zeta_0 g_{\omega}^2 (V_{\mu}V^{\mu})^2\nonumber\\&+&(1+\eta_{\rho} \frac{g_{\sigma} \sigma}{M})m_{\rho}^2 {\rm Tr}(\bar{b}_{\mu}\bar{b}^{\mu})-m_{\sigma}^2\sigma^2(1+\frac{\kappa_3}{3 !} \frac{g_{\sigma} \sigma}{M}+\frac{\kappa_4}{4 !} \frac{g_{\sigma}^2 \sigma^2}{M^2}),
\eea
where
\be
\bar{U}=\bar{\xi}^2,  ~ ~ ~ ~\bar{\nu}_{\mu \nu} = \partial_{\mu}\bar{\nu}_{\nu}-\partial_{\nu}\bar{\nu}_{\mu}+i [\bar{\nu}_{\mu},\bar{\nu}_{\nu}]=-i[\bar{a}_{\mu},\bar{a}_{\nu}].
\ee
In the mean field approximation, the $\pi$ meson does not have a contribution. 
If we set $\eta_1$,  $\eta_2$, $\zeta_0$, $\eta_{\rho}$ and $f_{\rho}$ equal to zero, in the Lagrangian density, the same equations of state for nucleons and mesons of the standard RMF models~\cite{pg,ring,serot} can be obtained.

The density dependence of the modified nuclear matter symmetry energy  after including the isovector-vector nonlinear term,
\be
{\mathcal L}_{\rm HP}= 4 \Lambda_V  g_{\rho}^2 g_{\omega}^2 ~\vec{b}^{\mu}\cdot \vec{b}_{\mu}~ V^{\mu}  V_{\mu},
\ee
 in the Lagrangian density of the standard RMF model and then followed by an adjustment of $ g_{\rho}$ and $\Lambda_V$ parameters has been for the first time calculated by Horowitz and Piekarewicz~\cite{Horowitz01}. Motivated by a similar philosophy, other additional nonlinear terms, but with different forms,  have also been studied in Ref.~\cite{Shen05}. In this paper, we follow the same procedure as given in Refs.~\cite{Horowitz01,Shen05}, but since we use  the E-RMF model which already contains an isovector-vector nonlinear term, the  density dependence of the nuclear matter symmetry energy can be modified without adding a new nonlinear term, instead, it only needs an adjustment of the  $g_{\rho}$ and  $\eta_{\rho}$ parameters while keeping the symmetry energy at the same value, i.e., $E_{\rm sym}$= 24.1 MeV at $k_F$ =1.14 fm$^{-1}$. The argument behind this procedure is that the symmetry energy at the saturation density ($k_F$ =1.32 fm$^{-1}$) is not well constrained experimentally. However, an average of symmetry energy at full density (the average density is less than saturation density) and at surface symmetry energy is constrained by binding energy of nuclei~\cite{Horowitz01,Shen05}. 

To study the effects of a $\delta$ meson  addition in the E-RMF model, we add to the Lagrangian density of that model the following terms, 
\be
{\mathcal L}_{\delta}=\frac{1}{2}(\partial_{\mu} \vec{\delta} \cdot \partial^{\mu} \vec{\delta}-m_{\delta}^2  \vec{\delta}^2)+\bar{\psi} g_{\delta} \vec{\tau} \cdot  \vec{\delta}\psi.
\ee
For electron and muon, the Lagrangian density reads
\be
\sum_{l=e^-,~\mu^-} \bar{l}( \gamma^{\mu}\partial_{\mu}-m_l)l.
\ee
All matter properties used in this work can be derived from these Lagrangian densities. A more detail step of  the derivations can be seen in Refs.~\cite{Furnstahl96, Wang00, sil,Horowitz01,Shen05,arumu}. Coupling constants for all parameter sets used in this work are displayed in Table~\ref{tab:params1}. To determine the fraction of every constituent, we imposed the requirement that in the neutron star at zero temperature, the chemical potential is in equilibrium and the charge is neutral.   
\begin{table}
\centering
\caption {Coupling constants of the parameter sets used in this work.}\label{tab:params1}
\begin{tabular}{crrrr}
\hline\hline Parameter &G2~~ & NL-3~~& NL-Z~ ~&Z271~~ \\\hline
$m_S/M$        &0.554~~& 0.541~~& 0.520~~ &0.495~~ \\
$g_S/(4 \pi)$ &0.835~~& 0.813~~& 0.801~~ &0.560~~ \\
$g_V/(4 \pi)$ &1.016~~& 1.024~~& 1.028~~ &0.670~~ \\
$g_R/(4 \pi)$ &0.755~~& 0.712~~& 0.771~~ &0.792~~ \\
$\kappa_3$     &3.247~~& 1.465~~& 2.084~~ &1.325~~ \\
$\kappa_4$     &0.632~~& -5.668~~& -8.804~~ &31.522~~ \\
$\zeta_0$      &2.642~~& 0~~& 0~~ &4.241~~ \\
$\eta_1$       &0.650~~& 0~~& 0~~ &0~~ \\
$\eta_2$       &0.110~~& 0~~& 0~~ &0~~ \\
$\eta_{\rho}$  &0.390~~& 0~~& 0~~ &0~~ \\\hline\hline
\end{tabular}\\
\end{table}

\subsection{Neutrino Mean Free Path}
For neutrino-electron matter interactions based on the standard model of weak interaction, the Lagrangian density for every constituent is
\be
{\mathcal L}_{\rm int}^j=\frac{G_F}{\sqrt{2}}
[\bar{\nu}\gamma^{\mu}(1-\gamma_5)\nu](\bar{\psi} J_{\mu}^j \psi),
\ee
where $J_{\mu}^j$ = $\gamma_{\mu}(C_V^j-C_A^j \gamma_5)$ and  ${\it j}$= $n, p, e^{-}, {\mu}^{-}$. The values of $C_V^j$ and $C_A^j$ can be seen in Table~\ref{tab:copconst0}.

\begin{table}
\centering
\caption {Coupling constants of neutrino-electron matter interactions. Here we use $\sin^2\theta_w=0.223$ and $g_A = 1.260$~\cite{reddy1,niembro01,horowitz91}.}\label{tab:copconst0}
\begin{tabular}{clr}
\hline\hline ~~Target~~ & ~~~~~~~~~~$C_V$~ ~ & $C_A~$ ~ ~  \\\hline\hline
$n$     &$-0.5$ & -$g_A$/2 ~ ~ \\
$p$     &$~~0.5-2\sin^2\theta_w$  & $g_A/2$ ~ ~ \\
$e$     &$~~0.5+2\sin^2\theta_w$  & $1/2$ ~ ~ \\
$\mu$ &$-0.5+2\sin^2\theta_w$~~~~  & $-1/2$  ~ ~ \\\hline\hline
\end{tabular}\\
\end{table}

The neutrino differential scattering cross-section can be derived from the Lagrangian density and has the form of  
 \begin{eqnarray}
\frac{1}{V}\frac{d^{3} \sigma}{d^{2}{\Omega}'dE'_{{\nu}}} &=& -\frac{G_{F}}{32{\pi}^{2}}\frac{E'_{{\nu}}}{E_{{\nu}}}{\rm Im}(L_{{\mu}{\nu}}{\Pi}^{{\mu}{\nu}}). 
\end{eqnarray}
Here $E_{{\nu}}$ and $ E'_{{\nu}}$ are the initial and final neutrino-electron energies, respectively,  $G_{F}= 1.023{\times} 10^{-5}/M^{2}$ is the weak coupling, and $M$ is the nucleon mass. The neutrino-electron tensor $L_{\mu\nu}$ can be written as 
\begin{eqnarray}
L_{{\mu}{\nu}} &=& 8[2k_{{\mu}}k_{{\nu}}+(k.q)g_{{\mu}{\nu}}-(k_{{\mu}}q_{{\nu}}+q_{{\mu}}k_{{\nu}})\nonumber\\
&-&{\it i}{\epsilon}_{{\mu}{\nu}{\alpha}{\beta}}k^{{\alpha}}q^{{\beta}}],
\end{eqnarray}
where $k$ is the initial neutrino-electron four-momentum and $q=(q_{0},{\vec{q}})$ is the four-momentum transfer. The polarization tensor ${\Pi}^{{\mu}{\nu}}$, which defines the target particles, can be written as 
\begin{eqnarray}
{\Pi}^{ j}_{{\mu}{\nu}}(q) &=& -i \int\frac{d^{4}p}{(2{\pi})^{4}}{\rm Tr}[G^{ j}(p)J^{ j}_{{\mu}}G^{ j}(p+q)J^{ j}_{{\nu}}],
\end{eqnarray}
where $p=(p_0,{\vec{p}})$ is the corresponding initial four-momentum, and $G(p)$ is the target particle propagator. The explicit form of the nucleon propagator is
\bea
G^{n,p}(p)&=&(\pslash^*+M^*)\Biggr[\frac{1}{p^*2-M^{*~2}+i\epsilon}+\frac{ i~\pi}{E^*}\nonumber\\&\times&\delta(p_0^*-E^*)\theta(p_F^{p,n}-\mid\vec{p}\mid)\Biggr],
\label{propagator}
\eea
where $E^*$ =  $E$ + $\Sigma_0$ is the nucleon effective energy and  $M^*$ = $E$ + $\Sigma_S$ is the nucleon effective mass. The  $\Sigma_0$ and $\Sigma_S$ are the scalar and time like self energies, respectively. Electron and muon propagators have similar expressions, except the effective (starred) quantities in Eq.~(\ref{propagator}) are replaced by the free ones.

The NMFP (symbolized by ${\lambda}$) as a function of the initial neutrino energy at a certain density is obtained by integrating the cross section over the time- and vector-component of the neutrino momentum transfer, as~\cite{horowitz91,niembro01} 
\begin{eqnarray}
\frac{1}{{\lambda}(E_{{\nu}})} = \int_{q_{0}}^{2E_{{\nu}}-q_{0}}d|{\vec{q}}|\int_{0}^{2E_{{\nu}}}dq_{0}\frac{|{\vec{q}}|}{E'_{{\nu}}E_{{\nu}}}
2{\pi}\frac{1}{V}\frac{d^3{\sigma}}{d^2{\Omega}'dE'_{{\nu}}}.\nonumber\\
\label{eq:nmfp}
\end{eqnarray}

Equation~(\ref{eq:nmfp}) is used to calculate the NMFP in the neutron star matter of Figs. \ref{nmfp}, \ref{mstrmfp}, and \ref{nmfpdel}.
\section{Results and Discussions}
\label{sec_rslt}
\begin{figure*}
\centering
 \mbox{\epsfig{file=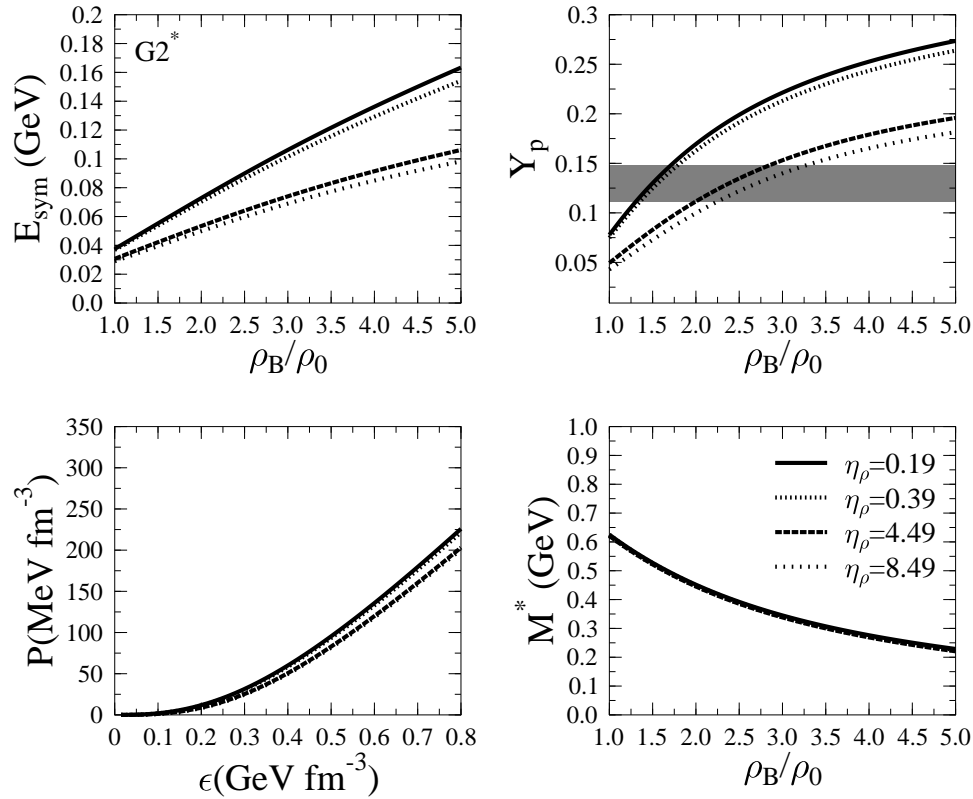,height=11.0cm}}
\caption{ Isovector-vector channel adjustment in the E-RMF model. Effects of different combinations of $g_{\rho}$ and $\eta_{\rho}$ on the symmetry energy of the nuclear matter are shown in the upper left panel, on the pressure and $M^*$ of the PNM are in the lower left and right panels, respectively, and on the proton fraction predictions in the upper right panel. Shaded region in the upper right panel corresponds to the proton fraction  threshold of the direct URCA process~\cite{lati}.\label{G2fac2}}
\end{figure*}

In this section we study the effects of the isovector-vector channel adjustment and the addition of $\delta$ meson in E-RMF models on the corresponding nuclear matter properties predictions. We also study  the role of every factor involved in the predicted NMFP in neutron stars of RMF models. We start with considering the effects of isovector-vector channel of the E-RMF model. Here we analyze the effects of  different g$_{\rho}$ and $\eta_{\rho}$ combinations of the G2 parameter set. The value of various coupling constant combinations can be seen in Table~\ref{tab:params2}. Their effects on matter properties are shown in Fig.~\ref{G2fac2}.

\begin{table}
\centering
\caption {Isovector-vector channel adjustment in the G2 parameter set of E-RMF model.}\label{tab:params2}
\begin{tabular}{crrrr}
\hline\hline \raisebox{-0.8ex}[0cm][0cm]{Isovector} & \multicolumn{4}{c}{Set} \\ \cline{2-5}
\raisebox{0.8ex}[0cm][0cm]{~~parameter~~} &I ~~~~~~~&II~~~~~~ &III~~~~~~ &IV~~~~~~ \\\hline
$g_{\rho}$     &9.358 ~ ~ & 9.483 ~ ~ & 11.786 ~ ~  &13.687 ~ ~ \\
$\eta_{\rho}$  &0.190 ~ ~ & 0.390 ~ ~ & 4.490 ~ ~ &8.490 ~ ~ \\\hline\hline
\end{tabular}\\
\end{table}

The symmetry energies of sets I--IV given in Table~\ref{tab:params2} are shown in the upper left panel. Since it has been pointed out by Lattimer $\it{et~al.}$~\cite{lati}, that the crucial role of the proton fraction  value for the onset of the direct URCA process is to enhance the neutron star cooling rate~\cite{baldo97}, we plot the corresponding proton fractions in the upper right panel. To estimate their effects on the EOS of neutron star, we plot the pressure as a function of energy of the pure neutron matter (PNM) as the dominant contributor in the EOS of a  neutron star, in the lower left panel. It is clear from the figure in the lower right panel that the adjustments of $g_{\rho}$ and $\eta_{\rho}$ have no effect on the PNM effective mass ($M^*$). 

In conclusion, for the E-RMF model, proton fraction predictions depend strongly on the behavior of the density dependent of $E_{\rm sym}$ at high density. On the other hand, the predicted neutron star EOS does not drastically depend on the behavior of $E_{\rm sym}$ at high density. 

\begin{table}
\centering
\caption{Isovector-vector channel adjustment in the Z271 parameter set of the Horowitz and Piekarewicz model~\cite{Horowitz01}.}\label{tab:params3}
\begin{tabular}{crrrr} 
\hline \hline \raisebox{-0.8ex}[0cm][0cm]{Isovector} & \multicolumn{4}{c}{Set} \\\cline{2-5}
\raisebox{0.8ex}[0cm][0cm]{~~parameter~~} &I ~~~~~~&II~~~~~~ &III~~~~~~ &IV~~~~~~ \\\hline
$g_{\rho}$     &9.498 ~ ~ & 9.672 ~ ~ & 11.506 ~ ~ &12.145 ~ ~ \\
$\Lambda_{V}$  &0 ~ ~ & 0.01 ~ ~ & 0.03 ~ ~ &0.035 ~ ~ \\\hline\hline
\end{tabular}\\
\end{table}

\begin{figure*}
\centering
 \mbox{\epsfig{file=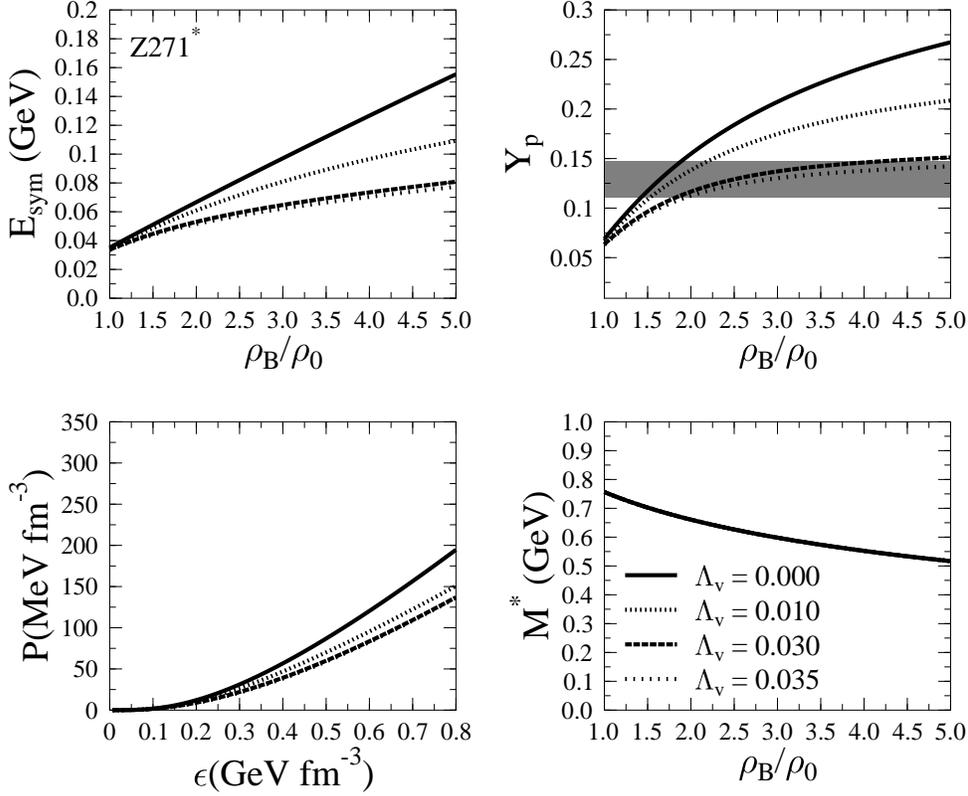,height=11.0cm}}
\caption{ Isovector-vector channel adjustment in the Horowitz and Piekarewicz model (Z271)~\cite{Horowitz01}. Effects of different combinations of $g_{\rho}$ and $\Lambda_{V}$ on the symmetry energy of SNM are shown in the upper left panel, on pressure and $M^*$ of PNM in the lower left and right panels, respectively, and on proton fraction predictions in the upper right panel. Shaded region in the upper right panel corresponds to   the proton fraction threshold for the direct URCA process~\cite{lati}.\label{eospiek}}
\end{figure*}

A similar analysis for the standard RMF plus an additional isovector-vector nonlinear term model of Horowitz and Piekarewicz~\cite{Horowitz01} with Z271* parameter set has been also performed. The various sets of coupling constant combinations are shown in Table~\ref{tab:params3}, whereas their effects on matter properties are shown in Fig~\ref{eospiek}. Similar conclusion is obtained both for proton fraction and $M^*$ in the PNM. Significant dependency in the EOS of this model on  $E_{\rm sym}$ is found. The different trend in  $E_{\rm sym}$ and EOS of this model compared to the E-RMF one originates from the different form of the isovector-vector non linear terms used in both models.
\begin{figure*}
\centering
 \mbox{\epsfig{file=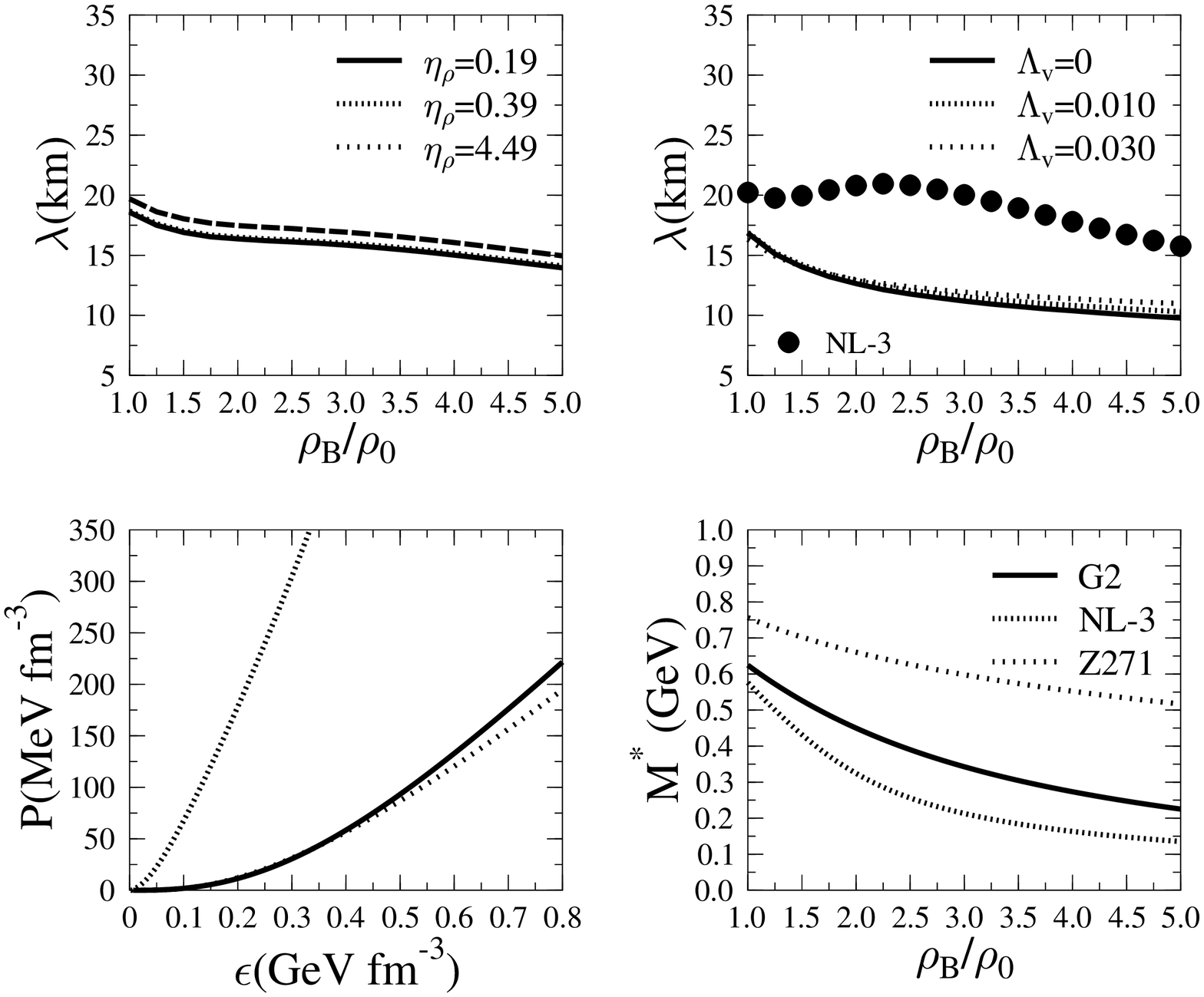,height=11.0cm}}
\caption{ Effects  of the isovector-vector channel adjustment on NMFP predictions for G2, NL-3 and Z271 models. The upper left panel is for the E-RMF model (G2) and the upper right panel is for the Horowitz and Piekarewicz model (Z271). For comparison, the result of NL-3 is given in the upper-right panel by the dots form. The pressure as a function of the energy density and $M^*$ as a function of the nucleon-saturation densities ratio are given in the lower left and lower right panels, respectively.\label{nmfp}}
\end{figure*}

To investigate which factor dominantly controls the behavior of the NMFP based on RMF models, we compare the EOS and NMFP of both models with the one of the standard RMF model, i.e., the NL-3 parameter set. The results are shown in Fig~\ref{nmfp}. In the upper left panel, we show the effects from the variation of $E_{\rm sym}$ in the G2* parameter sets (represented by the variation of  $g_{\rho}$ and $\eta_{\rho}$ values) on the NMFP. Similarly, for the Z271* and NL-3 parameter sets (represented by the dots form),  the results are shown in the upper right panel. The EOS and $M^*$ in the PNM of the standard G2, Z271 and NL-3 parameter sets are displayed in the lower left and right panels, respectively. Different from NL-3, which has an anomalous behavior in its NMFP, it is found that   G2* and  Z271* parameter sets predict the NMFP trends, which do not change with the variation of  $E_{\rm sym}$ (in both models, the anomalous behavior in the NMFP does not appear in every parameter set even though they have different E$_{\rm sym}$ values). This means that the appearance of the anomalous behavior in the NMFP seems to be insensitive to the value of the proton fraction. The NL-3 parameter set has PNM with a stiff EOS and a relatively small $M^*$ value but, on the contrary, Z271 and G2 have PNM with a soft EOS and large $M^*$ value at high density. This fact gives an indication that a soft EOS and a normal behavior of the NMFP are mostly determined by the relatively large $M^*$ value at high density. On the other hand, finite nuclei calculations using the standard RMF model~\cite{pg,ring,serot,anto05} inform us that acceptable shell structure predictions in finite nuclei regions require a small $M^*$ value ($\sim$ 0.6 $M$) in the saturation density which is fulfilled by G2~\cite{sil} and NL-3~\cite{pg,ring,serot,anto05} parameter sets.  

\begin{table}
\centering
\caption {$M^*$ variations in the NL-Z parameter set of the standard RMF model.~\cite{anto05}}\label{tab:params4}
\begin{tabular}{crrr}
\hline\hline Parameter &NL-Z~~ & P-070~~& P-080~~ \\\hline
$g_S/(4 \pi)$ &$0.801$~~& 0.673~~& 0.575~~ \\
$g_V/(4 \pi)$ &$1.028$~~& 0.817~~& 0.632~~ \\
$\kappa_3$     &$2.084$~~& 2.463~~& 2.899~~ \\
$\kappa_4$     &$-8.804$~~& $-7.595$~~& 12.779~~ \\\hline\hline
\end{tabular}\\
\end{table}

\begin{figure*}
\centering
 \mbox{\epsfig{file=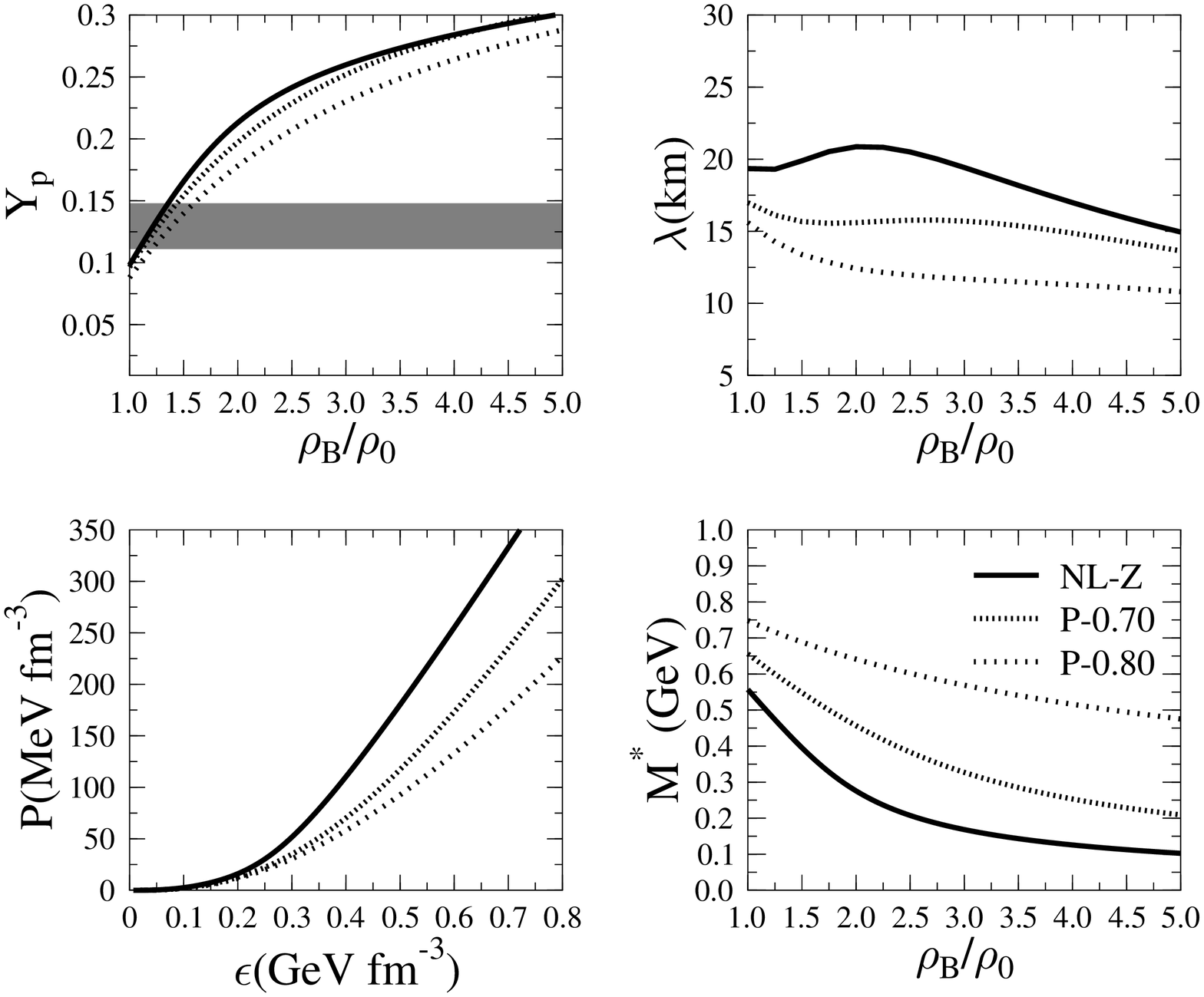,height=11.0cm}}
\caption{Effects of the $M^*$ variations on the proton fraction and on the  NMFP (upper left and upper right panels), and on the EOS (lower left panel).  Shaded region in the upper left panel corresponds to a threshold of the proton fraction for the direct URCA process~\cite{lati}.  Variations of the $M^*$ as a function of  the nucleon-saturation densities ratio is shown in the lower right panel.  \label{mstrmfp}}
\end{figure*}

Actually, the above results could be more clearly interpreted  by looking at the effects of different $M^*$ on the NMFP in the same model. The difference in each parameter set is only in the values of some adjusted coupling constants, while they should have acceptable SNM and PNM predictions at saturation density (more details about matter predictions of these models can be seen in Ref.~\cite{anto05}), and for our purpose here, the proton fraction predictions should be similar. The parameter sets of Ref.~\cite{anto05} are suitable for this task. The coupling constants variations of the models are shown in Table~\ref{tab:params4}, whereas the results are shown in Fig.~\ref{mstrmfp}. The effects of the $M^*$ variation on the proton fraction are shown in the upper left panel, while the effects on the NMFP  and EOS of PNM are shown in the upper right and lower left panels, respectively. Clearly, if $M^*$ becomes too low then the corresponding EOS becomes too stiff and the anomalous behavior in the NMFP appears. The P-070 and P-080 parameter sets have unacceptable shell structure predictions in some nuclei~\cite{anto05} since these parameter sets have a too large $M^*$ in the saturation density which, as a consequence, leads to a too narrow spin-orbit splitting prediction. Recently, ``FSU GOLD'' parameter set has been introduced by Todd-Rutel and Piekarewicz~\cite{todd} which yields a soft EOS, while still accurately reproducing experimental data of binding energies and charge radii of some magic nuclei and also centroid energies for breathing mode of $^{208}\rm Pb$ and $^{90}\rm Zr$. Unfortunately the shell structure prediction of this parameter set is not  reported in that paper. Therefore, before drawing any further conclusion, we should wait for their full calculation, including the predicted shell structure properties of some magic nuclei.

In conclusion, these results confirm previous findings~\cite{sil,arumu} about the wide range of applications of the E-RMF model. In our view, the reason comes from the fact that the E-RMF model has a relatively small $M^*$ ($\sim 0.6~M$) in saturation density (demanding feature for finite nuclei) but a relatively large $M^*$ at high density (demanding feature for the neutron star). Extra nonlinear and tensor terms of this model compared to the standard one seem to be the source of this behavior (see Table \ref{tab:params1}). From the possibility that the density dependent  $E_{\rm sym}$  can be adjusted, the claim that RMF models predict relatively lower threshold densities for the direct URCA process and this fact can be considered as a weak point of the models~\cite{blaschke,kolo,migdal}, now can be re-explored. A precise density dependent $E_{\rm sym}$ at high density experimentally determined and/or extracted from the properties of the neutron star are needed in this case.    
\begin{table}
\centering
\caption {Effects of the $\delta$ meson on the G2 parameter set of the E-RMF model. Case $\eta_{\rho}=0.39$.}\label{tab:params5}
\begin{tabular}{crrr}
\hline\hline  \raisebox{-0.8ex}[0cm][0cm]{Isovector} & \multicolumn{3}{c}{Set} \\\cline{2-4}
\raisebox{0.8ex}[0cm][0cm]{~~parameter~~} &I ~~~~~~~&II~~~~~~ &III~~~~~~ \\\hline
$g_{\rho}$ &9.483 ~ ~ ~& 12.313~ ~ ~& 15.937 ~ ~ ~ \\
$g_{\delta}$ &0 ~ ~ ~& 5.026~ ~ ~& 7.540~ ~ ~ \\\hline\hline
\end{tabular}\\
\end{table}

\begin{table}
\centering
\caption {Effects of the $\delta$ meson in the G2* parameter sets of the E-RMF model. Case $\eta_{\rho}=4.49$.}\label{tab:params6}
\begin{tabular}{crrr}
\hline\hline  \raisebox{-0.8ex}[0cm][0cm]{Isovector} & \multicolumn{3}{c}{Set} \\\cline{2-4}
\raisebox{0.8ex}[0cm][0cm]{~~parameter~~} &I ~~~~~~~&II~~~~~~ &III~~~~~~ \\\hline
$g_{\rho}$ &11.786 ~ ~& 15.304~ ~ ~& 18.784 ~ ~ ~\\
$g_{\delta}$ & 0 ~ ~& 5.026~ ~ ~& 7.540~ ~ ~ \\\hline\hline
\end{tabular}\\
\end{table}

\begin{figure*}
\centering
 \mbox{\epsfig{file=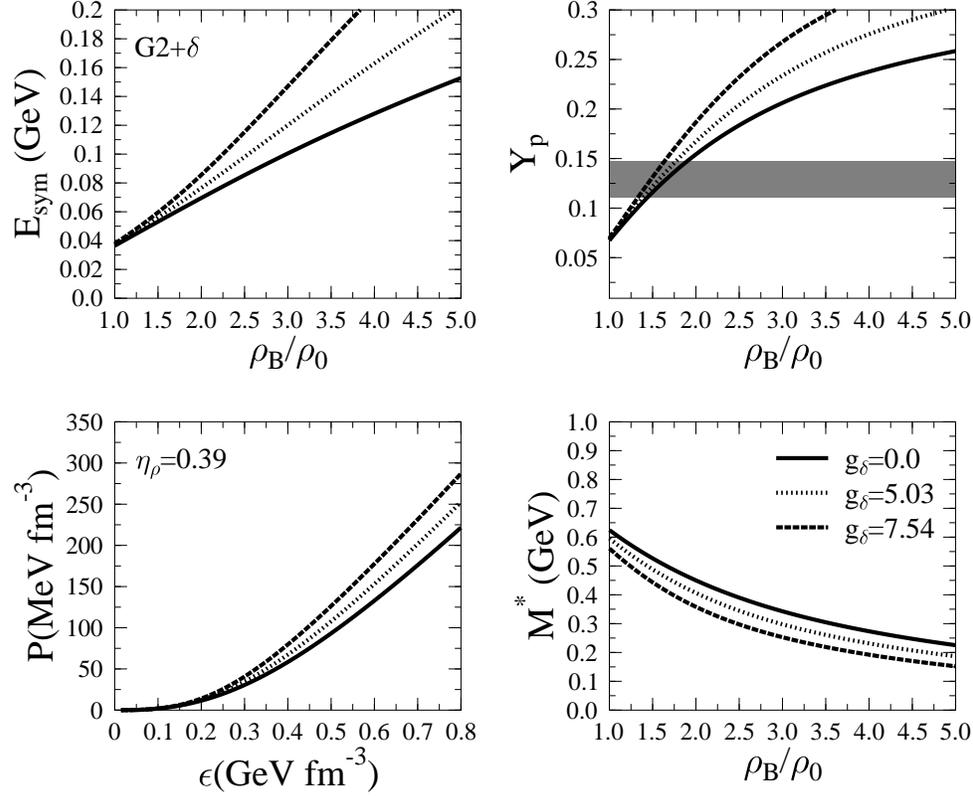,height=11.0cm}}
\caption{ Effects of the $\delta$ meson addition in the E-RMF model for $\eta_{\rho}=0.39$ on the symmetry energy and proton fraction (upper left and right panels) and on the EOS and $M^*$ in the PNM ( lower left and right panels).  Shaded region in the upper right panel corresponds to  the proton fraction  threshold for the direct URCA process~\cite{lati}.\label{delfrac}}
\end{figure*}

\begin{figure*}
\centering
 \mbox{\epsfig{file=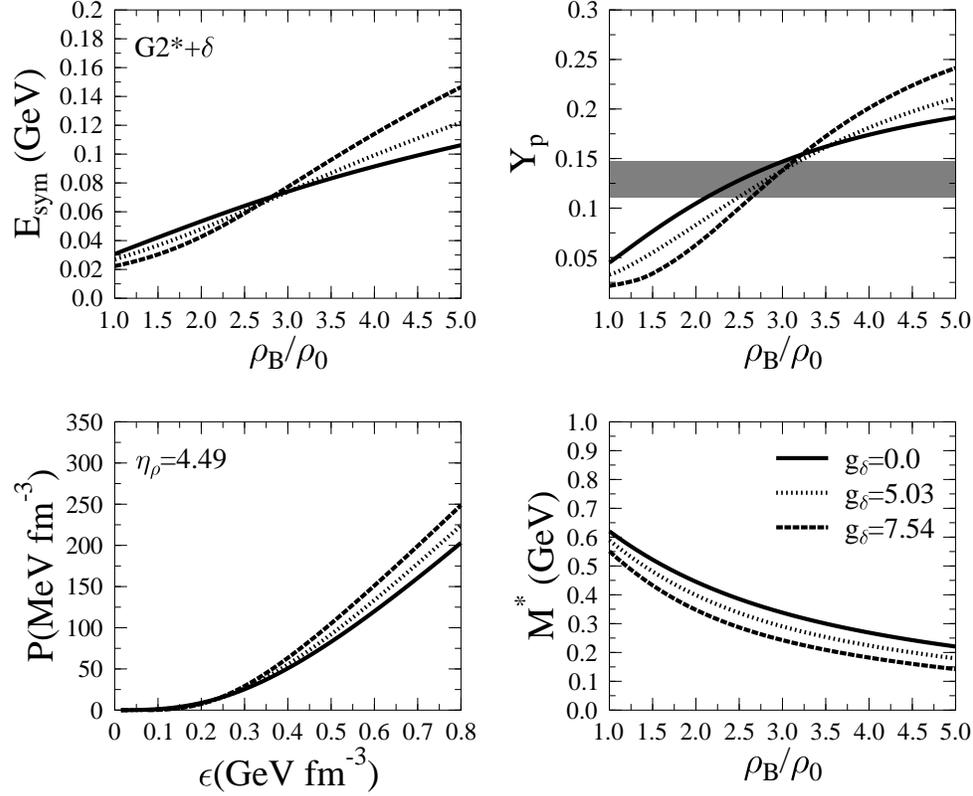,height=11.0cm}}
\caption{ Same as in Fig.~\ref{delfrac}, but for  $\eta_{\rho}=4.49$.\label{delfrac2}}
\end{figure*}

To study the  effects of $\delta$ meson on the E-RMF model, we start with the standard G2 with  $\eta_{\rho}=0.39$ and generate different  $g_{\rho}$ and $g_{\delta}$ parameters combinations but we keep the same $E_{\rm sym}=26.57$ MeV  at $k_F=1.172$ fm$^{-1}$ (note: there would be  no change in the conclusion if we took  $E_{\rm sym}=24.1$ MeV  at $k_F=1.14$ fm$^{-1}$). The combinations of the coupling constants of the models can be seen in Table~\ref{tab:params5}. The matter properties predictions are shown in Fig.~\ref{delfrac}. It is clearly seen that the presence of the $\delta$ meson results in a higher $E_{\rm sym}$ at high density. This fact leads to a higher proton fraction. But, in the region $\rho_B$=(1.5-2)$\rho_0$, the difference  between the $E_{\rm sym}$ value of the E-RMF plus a $\delta$ and that without a $\delta$ meson is not so significant. The presence of the $\delta$ meson also makes the  PNM EOS stiffer and reduces the value of the PNM $M^*$. The reduction magnitude depends on the magnitude of $g_{\delta}$. A similar trend is also found  in the  case of $\eta_{\rho}=4.49$. The combinations of $g_{\rho}$ and $g_{\delta}$ coupling constant are shown in Table~\ref{tab:params6} and the corresponding results can be seen in Fig.~\ref{delfrac2}.

\begin{figure*}
\centering
 \mbox{\epsfig{file=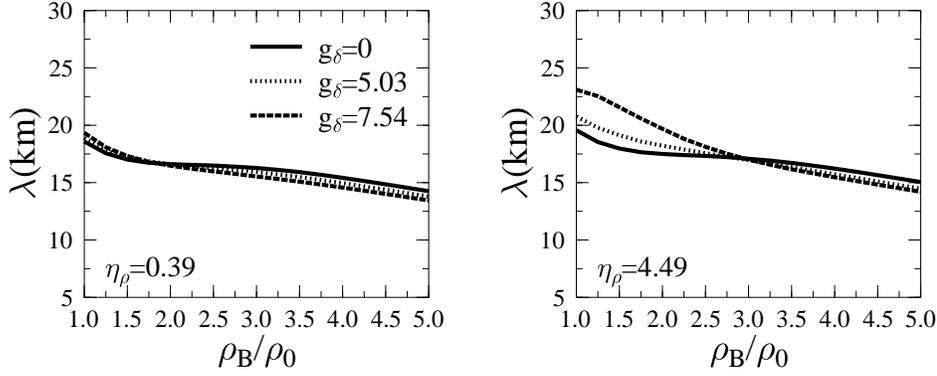,height=5.5cm}}
\caption{  Effects of the $\delta$ meson addition in the E-RMF model on the corresponding NMFP prediction. Left panel is for $\eta_{\rho}=0.39$, while right panel is for  $\eta_{\rho}=4.49$.\label{nmfpdel}}
\end{figure*}

In Fig.~\ref{nmfpdel}, it can be seen that the presence of the $\delta$ meson removes the anomalous behavior in the predicted NMFP. The effect appears more pronounce in the case  of $\eta_{\rho}=4.49$, rather than $\eta_{\rho}=0.39$. This fact is clearly depicted  in Fig.~\ref{nmfpdel}.

\begin{figure*}
\centering
 \mbox{\epsfig{file=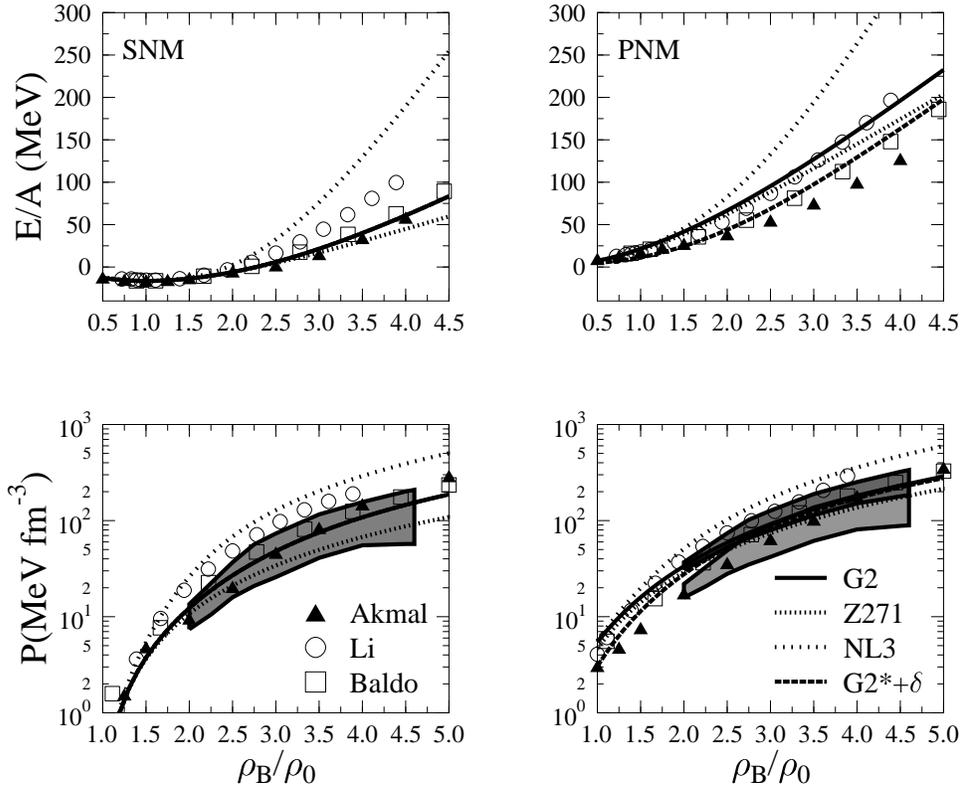,height=11.0cm}}
\caption{ Energy per nucleon ($E/A$) of the SNM (upper left) and PNM (upper right) of G2, Z271, NL3 and $G2^*+\delta$ parameter sets. The corresponding pressures are given in the lower left and lower right panels. For comparison, we also  show the results from variational calculation of Akmal {\it et al.}~\cite{akmal98}, DBHF calculation of Li {\it et al.}~\cite{li92}, and BHF with AV14 potential plus 3BF of Baldo {\it et ~al.}~\cite{baldo97}. Shaded regions correspond to experimental data from Danielewicz {\it et al.} \cite{daniel02}.\label{eos1}}
\end{figure*}

In Fig.~\ref{eos1}, we show the $E/A$ ratio and the pressure, for the SNM as well as the PNM of the four RMF parameter sets. NL-3 is a parameter set with good predictions for observables of finite nuclei and  has a stiff EOS at high density. Z271 is a parameter set that is specially constructed  for the neutron star and has a soft EOS at high density. G2 is a parameter set with acceptable predictions for observables of finite nuclei and has a relatively soft EOS at high density. Clearly, parameter sets with soft EOS are consistent with experimental data of Danielewicz $\it{et~al.}$~\cite{daniel02} and close to the results of the variational calculation by Akmal  $\it{et~al.}$~\cite{akmal98}, BHF calculation with AV14 potential plus 3BF of Baldo $et~al.$~\cite{baldo97}, and DBHF calculation of Li  $\it{et~al.}$~\cite{li92}. It seems also from the results of $G2^*+ \delta$, that the enhancement in the isovector channel of G2 shifts the $E/A$ ratio and pressure predictions of that parameter set closer to the result of variational calculation from Akmal  $\it{et~al.}$~\cite{akmal98}. 
\section{Summary}
In summary, we find that by adjusting the parameters in the isovector-vector sector of RMF models we can obtain a low proton fraction at high densities in neutron star. The anomalous behavior in the NMFP of RMF models would not appear, if their $M^*$ predictions at high density were sufficiently large. The presence of the $\delta$ meson in the E-RMF model increases the proton fraction at high density but the change in the value of the proton fraction is not significant in the region $\rho_B$=(1.5-2)$\rho_0$. For the PNM case, the presence of  the $\delta$ meson has effects that the EOS prediction becomes stiffer and $M^*$ becomes smaller at high density.  The presence of the $\delta$ meson in the E-RMF model also removes the anomalous behavior in the NMFP. By adjusting the isovector-vector sector and adding the $\delta$ meson in the E-RMF model, the $E/A$ ratio and the predicted pressure of the PNM become much closer to the results of  Akmal  $\it{et~al.}$~\cite{akmal98}.
\label{sec_sum}

\section*{ACKNOWLEDGMENT}
We acknowledge P. Danielewicz for his kindness to give us  his experimental data.

\end{document}